\documentclass[aps, prd, amsmath, floats, floatfix, twocolumn, nofootinbib, showpacs]{revtex4-1}
\usepackage{graphicx}
\usepackage{color}

\newcommand{\be}{\begin{eqnarray}}
\newcommand{\ee}{\end{eqnarray}}
\newcommand{\rar}{\rightarrow}

\begin{document}

\title{Attempt to explain black hole spin in X-ray binaries with new physics}

\author{Cosimo Bambi}
\email{bambi@fudan.edu.cn}

\affiliation{Center for Field Theory and Particle Physics and Department of Physics, 
Fudan University, 200433 Shanghai, China}

\date{\today}

\begin{abstract}
It is widely believed that the spin of black holes in X-ray binaries is mainly natal. A 
significant spin-up from accretion is not possible. If the secondary has a low mass, 
the black hole spin cannot change too much even if the black hole swallows the whole 
stellar companion. If the secondary has a high mass, its lifetime is too short to transfer 
the necessary amount of matter and spin the black hole up. However, while black 
holes formed from the collapse of a massive star with Solar metallicity are expected 
to have low birth spin, current spin measurements show that some black holes in X-ray 
binaries are rotating very rapidly. Here I show that, if these objects are not the Kerr 
black holes of general relativity, the accretion of a small amount of matter ($\sim 2$~$M_\odot$) 
can make them look like very fast-rotating Kerr black holes. Such a possibility is not 
in contradiction with any observation and it can explain current spin measurements 
in a very simple way.
\end{abstract}


\maketitle


\section{Introduction}

When a star exhausts all its nuclear fuel, it shrinks to find a new equilibrium 
configuration. For very massive stars, there is no known mechanism capable of 
balancing their own weight: these objects undergo a complete gravitational 
collapse and the final product is a black hole (BH). It is thought that in our Galaxy 
there are about $10^7$~BHs formed from the gravitational collapse of massive 
stars. Despite this huge number, we only know about 20~stellar-mass BH 
candidates~\cite{20}. They live in X-ray binaries and from the study of the orbital 
motion of the stellar companion it is possible to infer that the mass of the compact 
object exceeds 3~$M_\odot$. The latter is the maximum mass for a neutron or a 
quark star~\cite{rr74}, and therefore a compact object exceeding this limit is 
classified as a BH candidate.

In 4-dimensional general relativity, an uncharged BH is described by the Kerr 
solution and it is completely specified by only two parameters, corresponding 
to the mass $M$ and the spin angular momentum $J$ of the object. A fundamental 
limit for a Kerr BH is the bound $|a_*| \le 1$, where $a_* = J/M^2$ is the 
dimensionless spin parameter\footnote{Throughout the paper, I use units in 
which $G_{\rm N} = c = 1$.}. For $|a_*| > 1$ there is no event horizon in the Kerr 
metric and the spacetime has a naked singularity~\cite{bf09}. If we can measure 
both $M$ and $a_*$ of a Kerr BH, we know all the properties of the spacetime 
geometry. The effect of the accretion disk on the background metric is indeed 
negligible~\cite{disk}. However, it is not easy to estimate the BH spin: the spin 
has no effects in Newtonian gravity and therefore it is necessary to probe the 
spacetime close to the object. At present, the spin parameter has been 
measured only for about 10~stellar-mass BH candidates~\cite{spin-m}.

It is commonly thought that the spin of stellar-mass BHs in X-ray binaries is mainly 
natal and that the effect of the accretion process is negligible~\cite{kk99} (but see 
Ref.~\cite{fragos}). The argument can be summarized as follows. Stellar-mass BH 
candidates have a mass around 10~$M_\odot$. If the stellar companion is a few 
Solar masses, the BH cannot significantly change its mass and spin angular 
momentum even swallowing the whole star. If the stellar companion is heavy, 
its lifetime is too short: even if the BH accretes at the Eddington rate, there is not the 
time to transfer the necessary amount of matter to significantly spin the BH up.
In the end, a BH cannot swallow 
more than a few $M_\odot$ from the companion star, and for a $10$~$M_\odot$ 
object this is not enough to significantly changes its spin parameter $a_*$~\cite{kk99}.

BH binaries can be grouped into 2 classes. Low-mass X-ray binaries are systems 
in which the stellar companion is not more than a few Solar masses 
($\lesssim 3$~$M_\odot$) and the mass transfer occurs for Roche lobe overflow. 
These systems are transient X-ray sources because the mass transfer is not 
continuous. High-mass X-ray binaries are systems in which the stellar companion 
is massive ($\gtrsim 10$~$M_\odot$) and the mass transfer from the companion 
star to the BH is due to the wind of the former. These systems are persistent X-ray 
sources. If the BH spin is mainly natal, its value should be explained by studying 
the gravitational collapse of massive stars. While there are still uncertainties in 
the angular momentum transport mechanisms of the progenitors of stellar-mass BHs, 
it is widely accepted that the gravitational collapse of a massive star with Solar 
metallicity cannot create fast-rotating remnants~\cite{prog}. The birth spin of these 
BHs is expected to be very low (see e.g.~\cite{fragos} and references therein). 
However, this is not what we observe. Assuming the Kerr metric, 
BH spin measurements show that some of
these objects have a spin parameter close to 1. 
In the case of low-mass X-ray binaries, the BH candidate in GRS~1915+105 
has $a_* > 0.98$~\cite{grs1915} and $M = 12.4 \pm 2.0$~$M_\odot$~\cite{reid},
while the stellar companion's mass is $M = 0.52 \pm 0.41$~$M_\odot$. In the case
of high-mass X-ray binaries, the BH candidate in Cygnus~X-1 has
$a_* > 0.98$~\cite{cygx1} and $M = 14.8 \pm 1.0$~$M_\odot$, while the
stellar wind from the companion is not an efficient mechanism to transfer mass.
Both the spin constraints are at 3~$\sigma$. While BHs in low- and high-mass X-ray 
binaries form in different environments, in both cases the origin of so high spin 
values is puzzling: the birth spin is expected to be low and accretion can spin a 
BH up only by transferring a significant amount of matter.

In this paper, I show that current spin measurements can be easily explained if  
BH candidates in X-ray binaries are not the Kerr BHs of general relativity.
In particular, an initially non-rotating BH can look like a fast-rotating Kerr BH 
after accreting a small amount of matter ($\sim 2$~$M_\odot$) if it is more prolate 
than a Kerr BH. Strictly speaking, this does not necessary mean that the BH must 
be prolate, but simply that it must be less oblate than the Kerr one. Here the 
key-point is the innermost stable circular orbit (ISCO), which depends on the 
background metric. A BH more prolate than a Kerr one can look like a very 
fast-rotating Kerr BH when its spin parameter is much lower, which can be 
acquired after accreting a modest amount of mass. While the scenario is speculative 
and requires new physics, it is not in contradiction with any observation or theoretical 
argument~\cite{review}, and it provides a simple explanation to current spin 
measurements.

\section{Kerr Black Holes}

Accretion process from a thin disk is an efficient mechanism to spin a BH up. 
The inner edge of the disk is at the ISCO radius, as supported by observations~\cite{jack}. 
The accreting gas moves on nearly geodesic circular orbits on the equatorial 
plane. As the gas loses energy and angular momentum, it first approaches the 
ISCO radius and then quickly plunges onto the BH. The mass and spin angular 
momentum of the BH change as
\be
M \rar M + \delta M \, , \quad J \rar J + \delta J \, ,
\ee
where $\delta M$ and $\delta J$ are the mass and the angular momentum carried 
by the gas
\be
\delta M = E_{\rm ISCO} \delta m \, , \quad \delta J = L_{\rm ISCO} \delta m \, ,
\ee
$E_{\rm ISCO}$ and $L_{\rm ISCO}$ are, respectively, the specific energy and 
the specific angular momentum of the gas at the ISCO radius, while $\delta m$ 
is the gas rest-mass. With this set-up, one finds the well-known equation of the 
spin evolution~\cite{th74}
\be\label{eq-spin}
\frac{da_*}{d\ln M} = \frac{1}{M} 
\frac{L_{\rm ISCO}}{E_{\rm ISCO}} - 2 a_* \, .
\ee

In the case of the Kerr metric, it is possible to integrate Eq.~(\ref{eq-spin}) and 
find an analytic expression for the spin parameter $a_*$ as a function of the BH 
mass $M$~\cite{bardeen}
\be
\hspace{-0.3cm}
a_* =  
\begin{cases}
\sqrt{\frac{2}{3}}
\frac{M_0}{M} \left[4 - \sqrt{18\frac{M_0^2}{M^2} - 2}\right] 
 & \text{if } M \le \sqrt{6} M_0 \, , \\
1 & \text{if } M > \sqrt{6} M_0 \, ,
\end{cases}
\ee
assuming an initially non-rotating BH with mass $M_0$. The equilibrium value of 
the spin parameter is 1 and requires that the BH has increased its mass by a factor 
$\sqrt{6} \approx 2.4$. If we include the effect of the radiation emitted by the disk 
and captured by the BH, we find that the equilibrium spin parameter is about 0.998, 
the so-called Thorne limit~\cite{th74}, since radiation with angular momentum opposite 
to the BH spin has larger capture cross section.

\begin{figure*}
\begin{center}
\includegraphics[height=6cm]{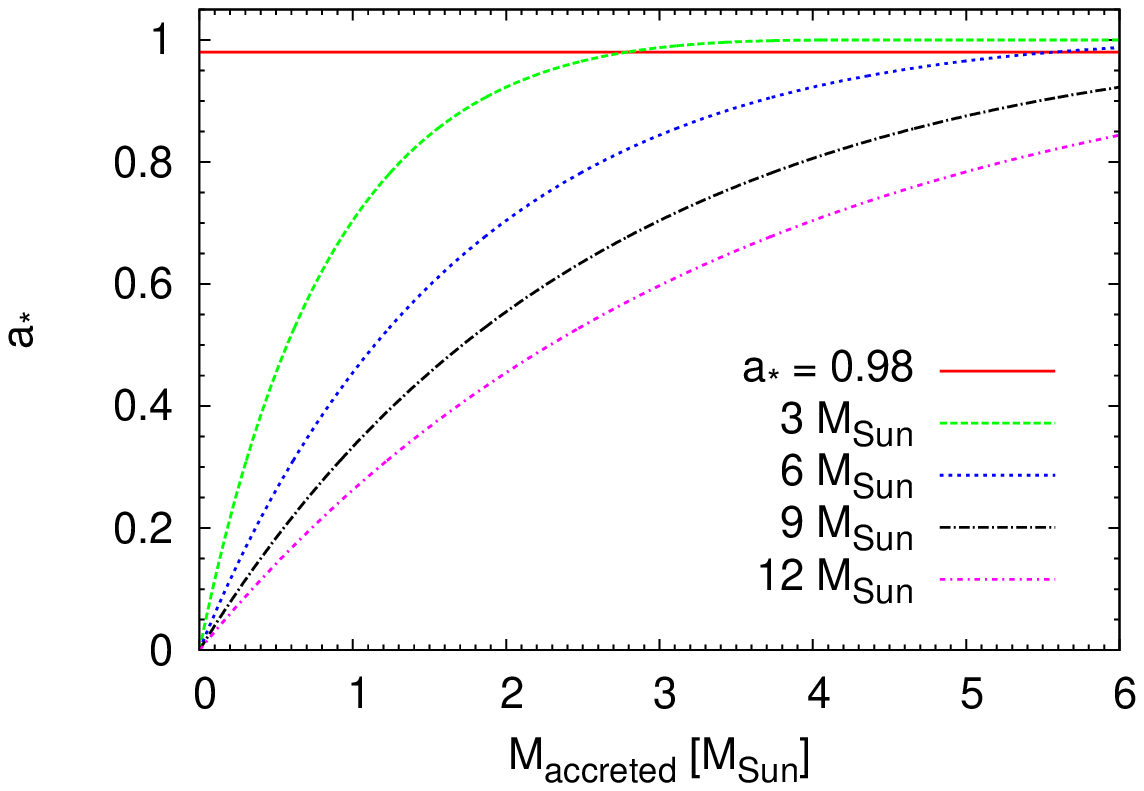}
\includegraphics[height=6cm]{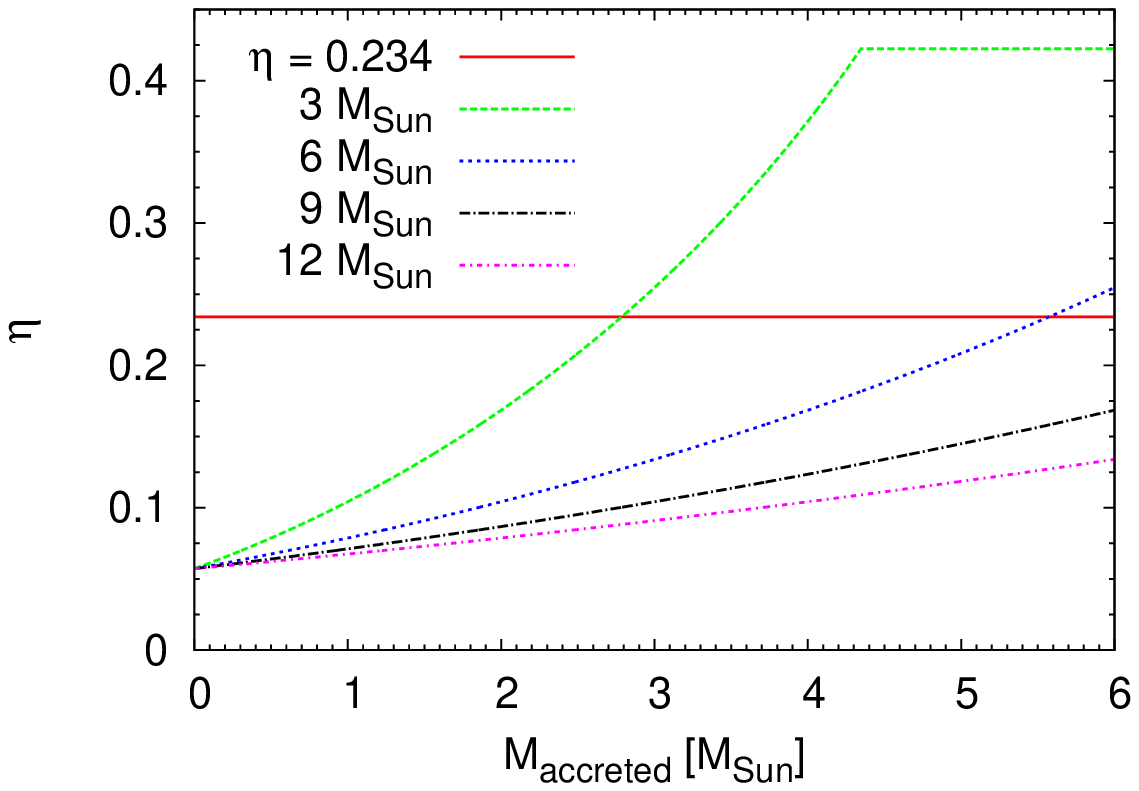}
\end{center}
\vspace{-0.5cm}
\caption{Kerr BHs. Evolution of the spin parameter $a_*$ (left panel) and of the 
radiative efficiency $\eta = 1 - E_{\rm ISCO}$ (right panel) as a function of the amount 
of matter accreted onto an initially non-rotating BH for four different initial BH masses 
(3~$M_\odot$, 6~$M_\odot$, 9~$M_\odot$, and 12~$M_\odot$). The horizontal 
red solid lines indicate the spin parameter $a_* = 0.98$ and the corresponding 
radiative efficiency $\eta = 0.234$. See the text for more details. \label{f1}}
\vspace{0.2cm}
\begin{center}
\includegraphics[height=6cm]{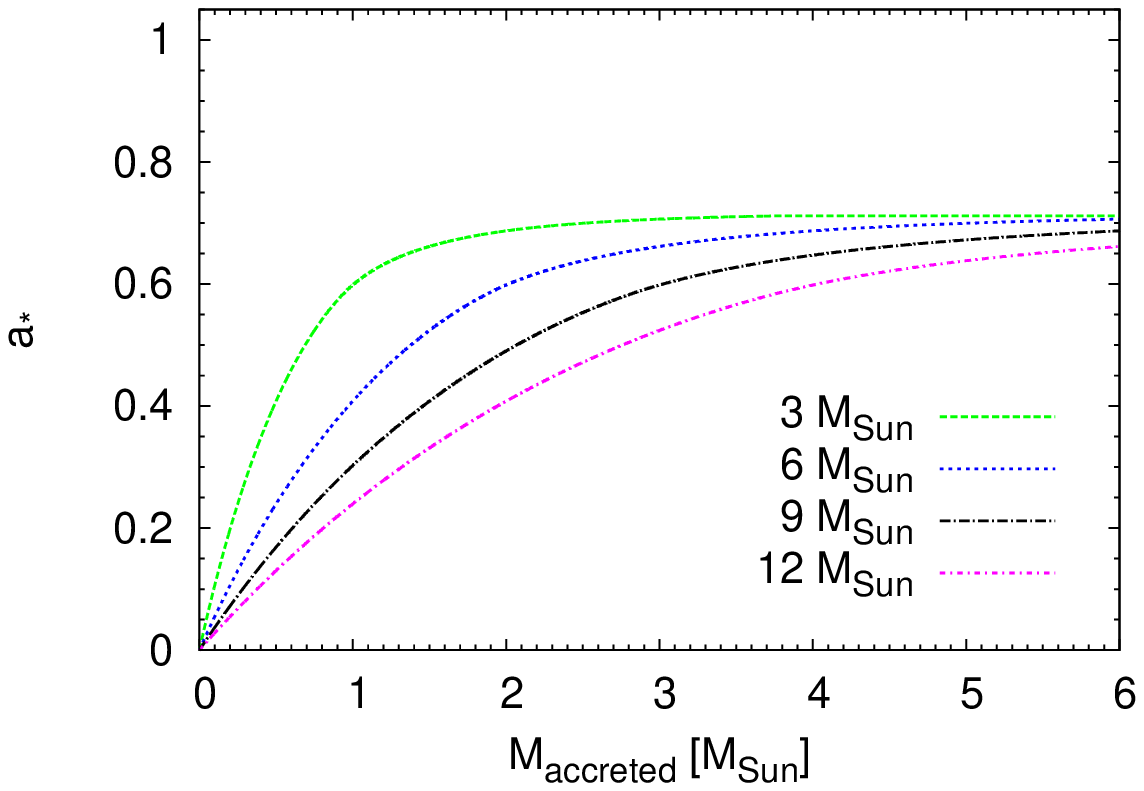}
\includegraphics[height=6cm]{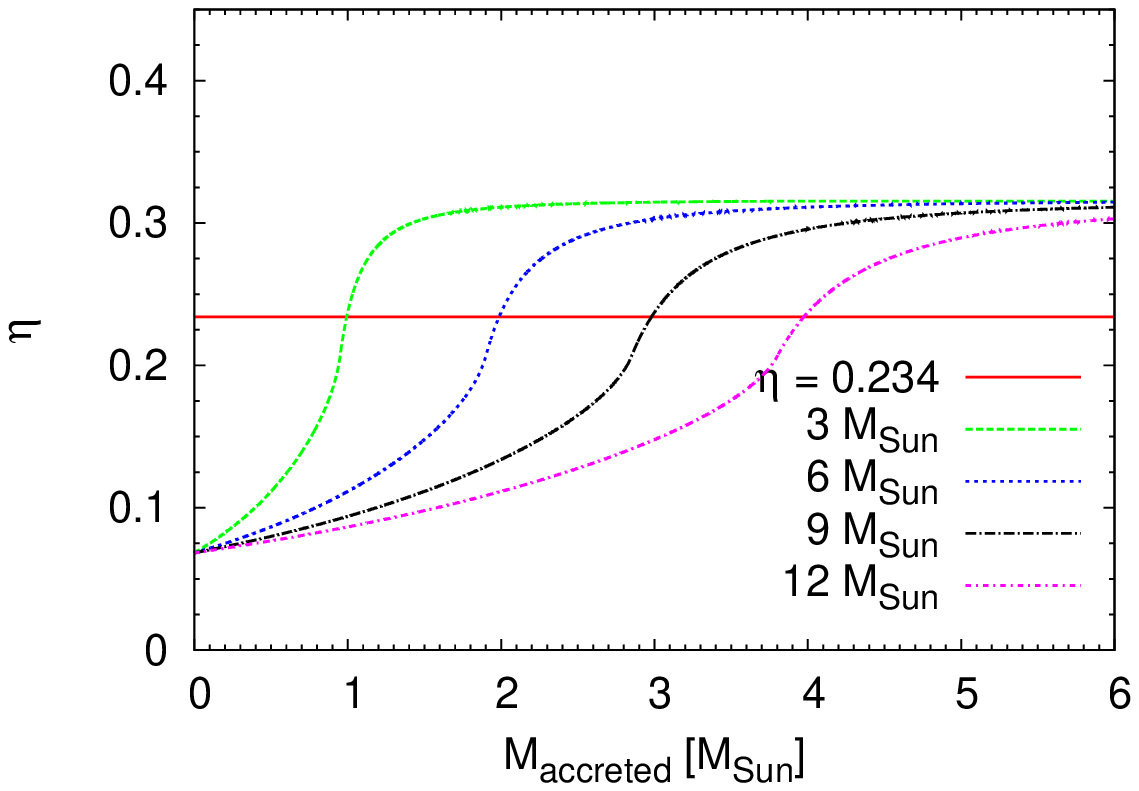}
\end{center}
\vspace{-0.5cm}
\caption{Non-Kerr BHs. As in Fig.~\ref{f1}, but in the case of Johannsen-Psaltis
BHs with $\epsilon_3 = 4$. The key-point is that here $\eta$ can become 
high after a modest mass transfer. See the text for more details. \label{f22}}
\vspace{0.2cm}
\begin{center}
\includegraphics[height=6cm]{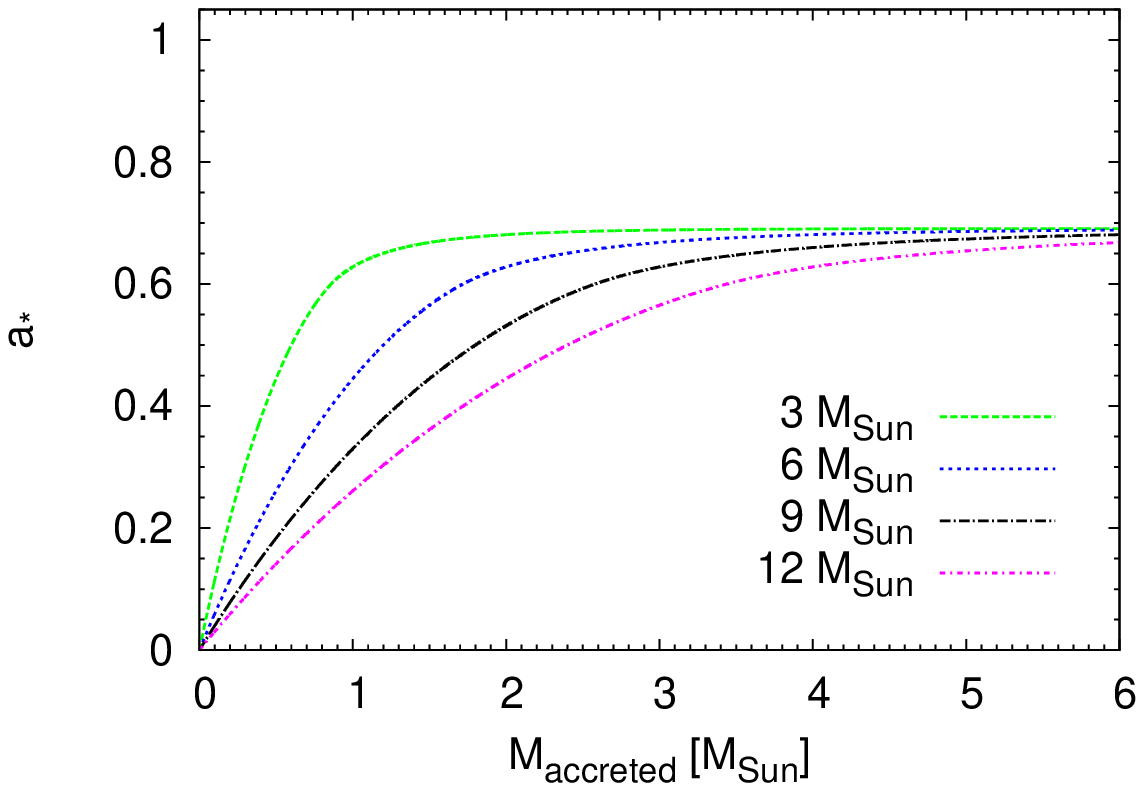}
\includegraphics[height=6cm]{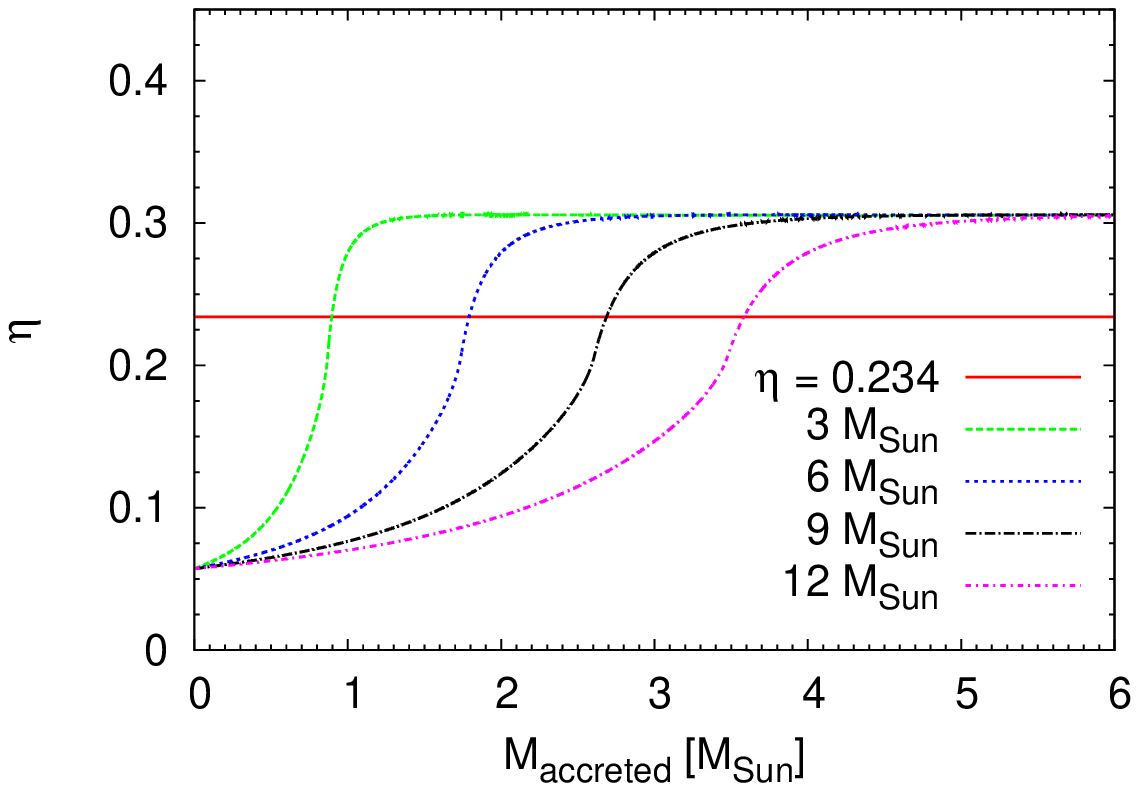}
\end{center}
\vspace{-0.5cm}
\caption{Non-Kerr BHs. As in Fig.~\ref{f1}, but in the case of Johannsen-Psaltis
BHs with $\epsilon_3 = 10 \, a_*^2$. See the text for more details. \label{f2}}
\end{figure*}

The left panel in Fig.~\ref{f1} shows the evolution of $a_*$ as a function of the accreted 
mass for some values of the initial BH mass. An initially non-rotating BH has to double 
its original mass to get $a_* = 0.98$. If a BH in an X-ray binary cannot strip more than 
a few Solar masses from the stellar companion, only for low mass BHs with 
$M \approx 3$~$M_\odot$ it may be possible to get $a_* = 0.98$. An initially low value of 
the spin parameter does not help very much, since the evolution of the spin 
parameter is faster at the beginning and slower when the spin is higher. As the birth
spin of BHs is expected to be low, we do not understand why we observe some 
fast-rotating BHs like GRS~1915+105. The latter should have been born with a mass 
of $\sim 6$~$M_\odot$ and have accreted a similar amount of matter from the stellar 
companion, which seems to be unlikely. The right panel in Fig.~\ref{f1} shows the 
evolution of the same systems in terms of the radiative efficiency $\eta = 1 - E_{\rm ISCO}$. 
A Kerr BH with $a_* = 0.98$ has a radiative efficiency $\eta = 0.234$.

\section{Non-Kerr Black Holes}

If BH candidates in X-ray binaries were not the Kerr BHs of general relativity, 
current spin measurements would be wrong (because obtained assuming the Kerr 
metric) and the evolution of the spin parameter as a result of mass transfer from 
the stellar companion would be different. A number of studies has clearly shown 
that there is a fundamental degeneracy between the spin and possible deviations 
from the Kerr solution, with the result that a non-Kerr BH may be interpreted as a 
Kerr BH with a different value of $a_*$~\cite{degeneracy}. The main technique
to estimate the spin parameter of stellar-mass BH candidates is the so-called
continuum-fitting method, namely the study of the thermal spectrum of thin 
disks~\cite{spin-m}. As a first crude approximation, the approach measures
the radiative efficiency $\eta$~\cite{lingyao}, which is then translated into a spin 
measurement under the assumption of the Kerr background, exploiting the fact 
there is a one-to-one correspondence between $\eta$ and $a_*$. It turns out
that BHs more prolate than the Kerr ones have a higher radiative efficiency for a 
lower value of $a_*$ and they can thus look like very fast-rotating Kerr BHs 
after acquiring a relatively small amount of mass from the stellar-companion. 
The result is very general, but it is useful to see this with some specific example.

As first case, I consider the Johannsen-Psaltis metric~\cite{jp}. 
Observational constraints on this metric from stellar-mass BH candidates are
discussed in~\cite{lingyao}. 
In Boyer-Lindquist coordinates, the line element reads~\cite{jp}
\begin{widetext}
\be\label{eq-jp}
ds^2 &=& - \left(1 - \frac{2 M r}{\Sigma}\right) (1 + h) \, dt^2
+ \frac{\Sigma (1 + h)}{\Delta + a^2 h \sin^2\theta } \, dr^2
+ \Sigma \, d\theta^2 - \frac{4 a M r \sin^2\theta}{\Sigma} (1 + h) \, dt \, d\phi + \nonumber\\
&& + \left[\sin^2\theta \left(r^2 + a^2 + \frac{2 a^2 M r \sin^2\theta}{\Sigma} \right)
+ \frac{a^2 (\Sigma + 2 M r) \sin^4\theta}{\Sigma} h \right] d\phi^2 \, ,
\ee
\end{widetext}
where $a = J/M $, $\Sigma = r^2 + a^2 \cos^2\theta$, and $\Delta = r^2 - 2 M r + a^2$.
$h$ introduces deviations from the Kerr background and in its simplest version it is given 
by
\be
h = \frac{\epsilon_3 r M^3}{\Sigma^2} \, ,
\ee
where $\epsilon_3$ is an unknown parameter that quantifies possible deviations from 
the Kerr solution. Johannsen-Psaltis BHs are more prolate (oblate) than their 
Kerr cousins with the same spin parameter when $\epsilon_3 > 0$ 
($\epsilon_3 < 0$)~\cite{topo}.

The Johannsen-Psaltis metric is a phenomenological metric and does 
not describe any known solution of spinning BHs in modified gravity. It can be used 
as a toy-model to described non-Kerr BHs assuming that particles follow the geodesics 
of its spacetime, namely that it can be obtained as a solution (or approximated solution) 
of some metric theory of gravity. In this case, it is possible to study 
the evolution of the spin parameter of these non-Kerr objects~\cite{spin-cb}. The master 
equation is still~(\ref{eq-spin}), but $E_{\rm ISCO}$ and $L_{\rm ISCO}$ are different because 
they are determined by the background metric.

In the original proposal of Ref.~\cite{jp}, $\epsilon_3$ is a 
phenomenological parameter quantifying possible deviations from the Kerr geometry. 
In presence of an underlying theory, it may be related to some coupling parameter in 
the modified action and thus be a constant of the theory. Fig.~\ref{f22} shows the 
evolution of the spin parameter $a_*$ and of the radiative efficiency $\eta$ of these 
Johannsen-Psaltis BHs for $\epsilon_3 = 4$. Another possible scenario is that
non-rotating BHs are spherically symmetric and that rotation makes the object more
and more deformed. For instance, Kerr BHs have mass-quadrupole moment given
by $Q = - a_*^2 M^3$: $Q=0$ in the non-rotating case and the BH is more and more 
oblate as the spin parameter increases. The mass-quadrupole moment of neutron 
stars can be approximated by $Q = - (1 + \chi) a_*^2 M^3$, where $\chi \sim 1-10$ 
is a parameter that mainly depends on the matter equation of states and at some 
level on the mass $M$~\cite{ns-chi}. Since $\chi > 0$, neutron stars are more oblate 
than Kerr BHs with the same spin parameter $a_*$. It is thus possible that $\epsilon_3$ 
depends on the spin parameter, and the simplest case is that it 
is proportional 
to $a_*^2$, say $\epsilon_3 = k a_*^2$, Fig.~\ref{f2} shows the evolution of the spin 
parameter $a_*$ and of the radiative efficiency $\eta$ of these Johannsen-Psaltis BHs
for $k = 10$. In both the scenarios, we find 
two remarkable features of these BHs: $i)$ the equilibrium
spin parameter is much lower than 1, and $ii)$ an initially non-rotating BH reaches
a high radiative efficiency very quickly, after a modest amount of mass transfer.
This is a preliminary indication that mass accretion onto a non-Kerr BH may explain 
the observation of fast-rotating Kerr BHs in X-ray binaries.

\begin{figure*}
\begin{center}
\includegraphics[height=6cm]{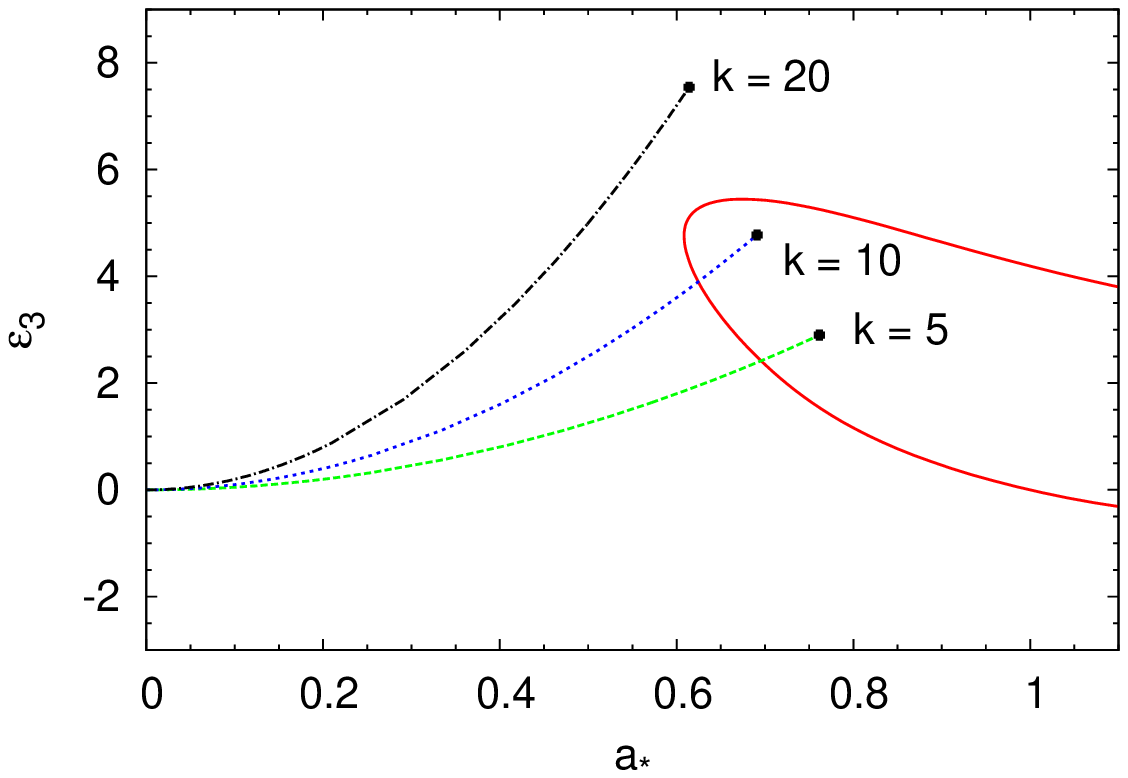}
\includegraphics[height=6cm]{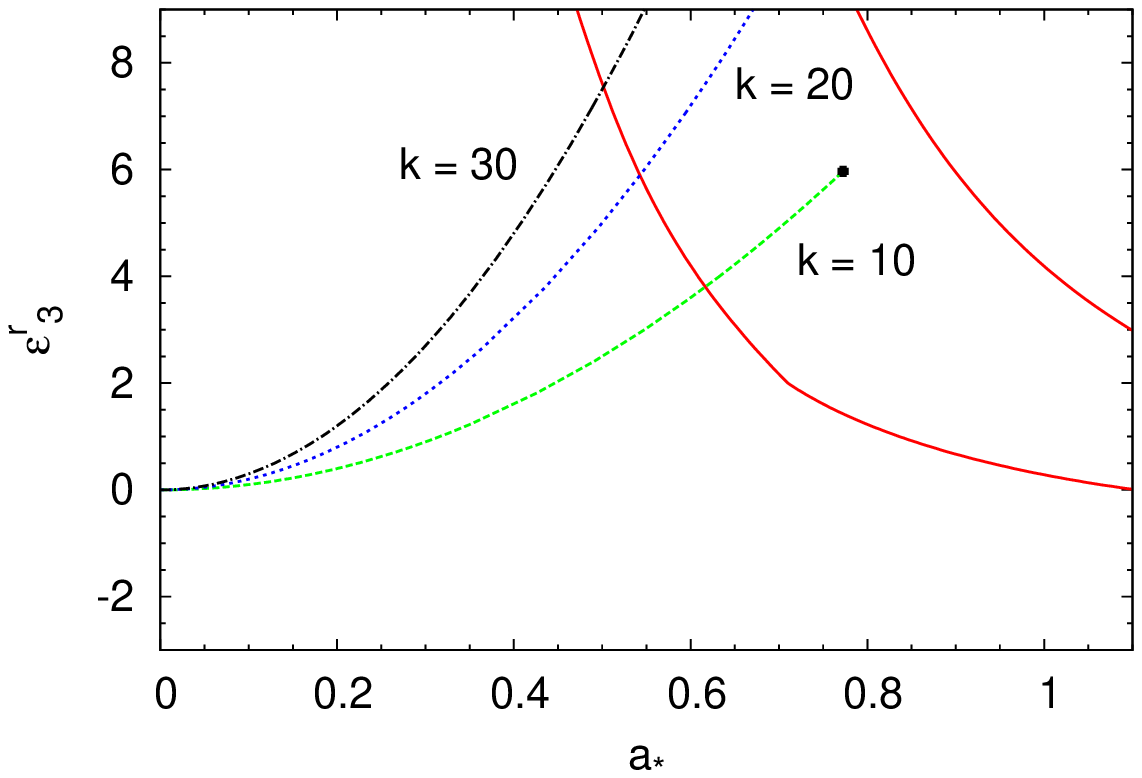}
\end{center}
\vspace{-0.5cm}
\caption{Left panel: evolution of Johannsen-Psaltis BHs with $\epsilon_3 = k a_*^2$
on the plane $(a_*, \epsilon_3)$ as a result of the accretion process. The black dots
indicate the equilibrium configurations. The red solid line is the boundary of the allowed 
region of the BH candidate in GRS~1915+105 (see Ref.~\cite{lingyao}). For $k=10$,
an initially non-rotating BH with a mass $M = 9.4$~$M_\odot$ enters the allowed 
region after a mass transfer $M_{\rm accreted} = 3.0$~$M_\odot$. Right panel: evolution
of Cardoso-Pani-Rico BHs with $\epsilon_3^r = k a_*^2$ on the plane $(a_*, \epsilon_3^r)$ 
as a result of the accretion process. The red solid line is the boundary of the allowed 
region of the BH candidate in Cygnus~X-1 (see Ref.~\cite{cpr-cb}). For $k=30$, an 
initially non-rotating BH with a mass $M = 12.4$~$M_\odot$ enters the allowed 
region after a mass transfer $M_{\rm accreted} = 2.4$~$M_\odot$. See the text for 
more details. \label{f3}}
\end{figure*}

The argument of the radiative efficiency can be used only for a preliminary estimate 
and deviations are more important for high values of $\eta$, mainly because in the 
spin measurement from the disk's thermal spectrum there is a correlation between 
the estimate of $a_*$ and the mass accretion rate~\cite{lingyao}. Ref.~\cite{lingyao} 
reports the current constraints on $a_*$ and $\epsilon_3$ of the 10~stellar-mass BH 
candidates with a spin measurement. The left panel in Fig.~\ref{f3} shows the evolution 
of some Johannsen-Psaltis BHs on the plane $(a_*,\epsilon_3)$ as well as the the 
boundary of the allowed region (red solid line) for the BH in GRS~1915+105 (the 
allowed region is the area inside the line). With the ansatz $\epsilon_3 = k a_*^2$, 
we can see that $k$ cannot be too high. For $k = 0$, we recover the standard Kerr 
metric and it is necessary a too large amount of matter to spin a BH up. For $k = 10$, 
an initially non-rotating BH with a mass $M = 9.4$~$M_\odot$ enters the allowed 
region after accreting 3.0~$M_\odot$. The final mass of 12.4~$M_\odot$ would 
correspond to the mass measurement of the BH in GRS~1915+105. For $k = 5$, 
an initially non-rotating BH with a mass $M = 9.1$~$M_\odot$ enters the allowed 
region after accreting 3.3~$M_\odot$. In the case of the BH in Cygnus~X-1, we can 
obtain similar results.

The case of the Johannsen-Psaltis BHs is just an 
example. The result is very general, in the sense that BHs more prolate than the 
Kerr ones can look like fast-rotating Kerr BHs after accreting a modest amount 
of mass. The necessary amount of mass is clearly related to the specific 
non-Kerr model. For instance, the Cardoso-Pani-Rico 
BHs~\cite{cpr} are a simple generalization of the Johannsen-Psaltis ones. In
this case, there is no systematic study of current constraints for all the BH
candidates, but the case of Cygnus~X-1 was analyzed in~\cite{cpr-cb}. 
In Boyer-Lindquist coordinates, the line element reads~\cite{cpr}
\begin{widetext}
\be\label{eq-m}
ds^2 &=& - \left(1 - \frac{2 M r}{\Sigma}\right)\left(1 + h^t\right) dt^2
- 2 a \sin^2\theta \left[\sqrt{\left(1 + h^t\right)\left(1 + h^r\right)} 
- \left(1 - \frac{2 M r}{\Sigma}\right)\left(1 + h^t\right)\right] dt d\phi \nonumber\\ 
&& + \frac{\Sigma \left(1 + h^r\right)}{\Delta + h^r a^2 \sin^2\theta} dr^2
+ \Sigma d\theta^2 
+ \sin^2\theta \left\{\Sigma + a^2 \sin^2\theta \left[ 2 \sqrt{\left(1 + h^t\right)
\left(1 + h^r\right)} - \left(1 - \frac{2 M r}{\Sigma}\right)
\left(1 + h^t\right)\right]\right\} d\phi^2 \, , \quad
\ee
\end{widetext}
where, in the simplest version, $h^t$ and $h^r$ are
\be
h^t = \epsilon_{3}^t \frac{r M^3}{\Sigma^2} \, , 
\quad
h^r = \epsilon_{3}^r \frac{r M^3}{\Sigma^2} \, .
\ee
As shown in Ref.~\cite{cpr-cb}, current observations can strongly constrain
$\epsilon_3^t$, while the bounds are very weak for $\epsilon_3^r$. We can thus
consider the case $\epsilon_3^t = 0$ and $\epsilon_3^r = k a_*^2$ and
compute the amount of mass transfer necessary to explain the BH spin in
Cygnus~X-1. The calculations are reported in the right panel in Fig.~\ref{f3}.
There are three cases, respectively for $k = 10$, 20, and 30. The area between
the two red solid curves is the region allowed by observations for Cygnus~X-1
and obtained in Ref.~\cite{cpr-cb}. The mass of the BH in Cygnus~X-1 is
$M = 14.8 \pm 1.0$~$M_\odot$. For $k = 30$, an initially non-rotating BH
with a mass $M = 12.4$~$M_\odot$ enters the allowed region after a mass 
transfer $M_{\rm accreted} = 2.4$~$M_\odot$. For a lower value of $k$ it
is necessary a larger amount of mass. For instance, in the case $k = 10$
an initially non-rotating BH with a mass $M = 11.6$~$M_\odot$ enters the 
allowed region after a mass transfer $M_{\rm accreted} = 3.2$~$M_\odot$.

It should not be difficult to find more efficient non-Kerr models, namely BHs
that can look like very-fast rotating Kerr BHs after accreting a smaller amount 
of matter. However, it has to be noted that the Johannsen-Psaltis and 
Cardoso-Pani-Rico spacetimes are phenomenological metrics to parametrize
generic deviations from the Kerr solution. Some spinning BH 
solutions in well-motivated alternative theories of gravity are known~\cite{q-grav}.
Generally speaking, if we consider a specific alternative theory of gravity, it is
possible that its BH solutions are not sufficiently more prolate than Kerr for any
choice of the parameters of the theory. Moreover, the deformation parameters
in the Johannsen-Psaltis and Cardoso-Pani-Rico spacetimes are only constrained 
by observations sensitive to the metric (assuming geodesic motion), not by the 
field equations of the theory (which is not given). If we have a theory, there may
be independent bounds coming from the field equations (e.g. emission of 
gravitational waves from a binary pulsar). After satisfying these constraints, it is
not obvious that their BHs can do the job proposed in the present paper. Every theory 
is different and it should be analyzed by itself. There are also scenarios like 
the one suggested in~\cite{g-cond}, in which general relativity holds up to quite
strong gravitational fields, but BHs have not the usual properties of BHs 
(even the concept of metric breaks down on the surface of these objects) 
and, in many aspects, they behave like compact stars made of exotic matter. In 
the latter case, constraints can only be obtained by BH observational data and
the current bounds on possible deviations from the Kerr solution are 
weak~\cite{lingyao}.

\section{Concluding remarks}

In this paper, I showed that current spin measurements of BHs in X-ray binaries may 
be easily explained if these objects are more prolate than the predictions of general 
relativity. The point is that similar objects can look like very fast-rotating Kerr BHs 
with a lower value of the spin parameter, which can be acquired after a modest mass 
transfer from the stellar companion. In the case of Kerr BHs, the required amount of 
mass stripped from the stellar companion is too high and therefore it is not understood 
the origin of so high spins for some BHs in X-ray binaries. It is at least intriguing that 
even other observations seem to require that stellar-mass BH candidates are more 
prolate than the Kerr ones, namely the power of steady jets~\cite{jet} and quasi-periodic
oscillations~\cite{qpo}. At this stage, the proposal that BH candidates are not the the 
Kerr BHs of general relativity is a very speculative possibility, but it is not in contradiction 
with any observation. It is probably difficult to test this scenario with a more detailed 
study of the origin of the BH spin, but future observational facilities will be 
hopefully able to test the Kerr nature of astrophysical BH candidates~\cite{future}.

\vspace{0.5cm}

{\it Acknowledgments ---}
This work was supported by the NSFC grant No.~11305038, 
the Shanghai Municipal Education Commission grant for Innovative 
Programs No.~14ZZ001, the Thousand Young Talents Program, 
and Fudan University.


\end{document}